\documentclass[
  aps,
  prl,
  reprint,
  nofootinbib,
  nobibnotes,
  floatfix
]{revtex4-2}

\usepackage[T1]{fontenc}
\usepackage[utf8]{inputenc}
\usepackage{lmodern}
\usepackage{microtype}
\usepackage{graphicx}
\usepackage{xcolor}
\usepackage{hyperref}
\usepackage{siunitx}

\setcitestyle{super,sort&compress}

\begin{document}

\title{MorpheusAI: agents autonomously generate, execute and refine virtual tissue models}

\author{Prerana Chandratre}
\author{Anjali Sharma}
\author{Justin Bürger Michaelis}
\author{Diego Jahn}
\author{Jörn Starruß}
\author{Lutz Brusch}
\thanks{These authors contributed equally to this work.}

\author{Simon Syga}
\thanks{These authors contributed equally to this work.}
\collaboration{Corresponding authors: \href{mailto:lutz.brusch@tu-dresden.de}{lutz.brusch@tu-dresden.de}, \href{mailto:simon.syga@tu-dresden.de}{simon.syga@tu-dresden.de}}
\affiliation{Center for Interdisciplinary Digital Sciences, TUD Dresden University of Technology, Dresden, Germany}

\begin{abstract}
Virtual tissue models link observations to mechanistic hypotheses, but building them demands expertise in biology, mathematics and programming.
Large language model agents using the simulator Morpheus (MorpheusAI) autonomously built such models for 15 benchmark studies, five without a published model.
With a frontier model, agents running simulations met 83--86\% of study-specific criteria, averaging under two hours and US\$4 per model.
Such agents could make \textit{in silico} modeling routine.
\end{abstract}

\maketitle

\begin{figure*}[t]
\centering
\includegraphics[]{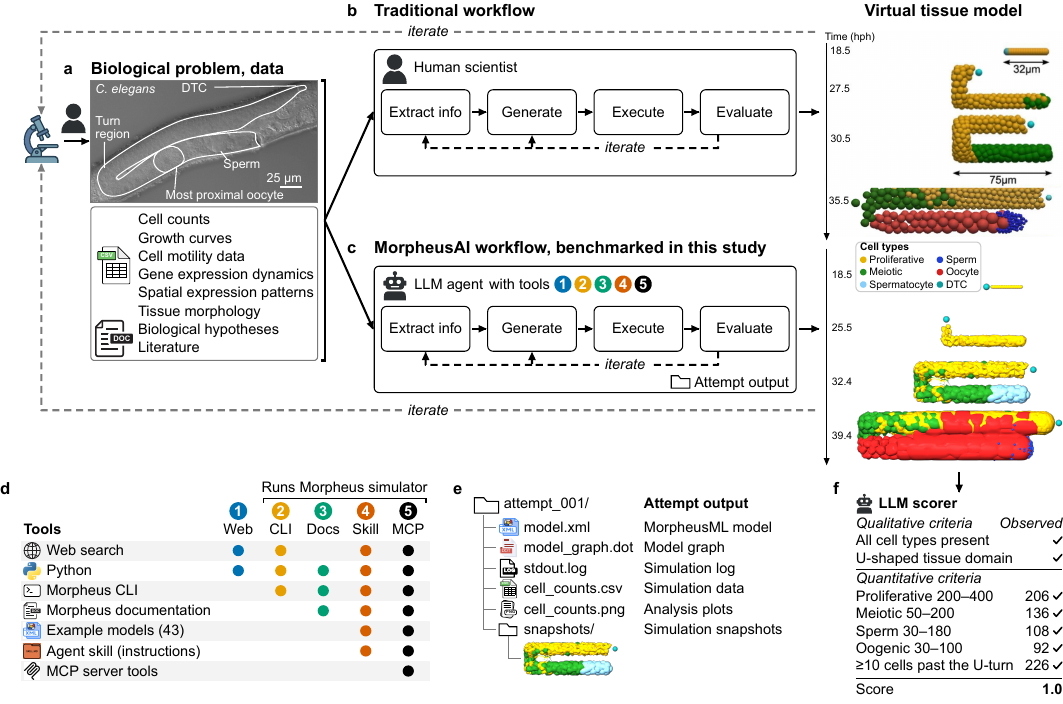}
\caption{\textbf{Autonomous construction of virtual tissue models.}
\textbf{a}, Biological problems and associated data, illustrated by germ-line development in \textit{C. elegans} (micrograph adapted from Fig. 1D of Atwell \textit{et al.}\cite{atwell2015}).
\textbf{b}, Traditional workflow: a human scientist extracts information from the observations, generates a virtual tissue model in software, executes it and evaluates the output, iterating until the model matches the data (simulation snapshots adapted from Fig. 3 of Atwell \textit{et al.}\cite{atwell2015}).
\textbf{c}, Agentic workflow benchmarked in this study: an LLM agent equipped with tools (numbered discs, the five setups) runs the same modeling loop without human intervention. Snapshots show the germ-line model built by GPT-6.1 Sol with setup 4 (Skill). The outer loop, in which insights from the model feed back into new experiments (gray dashed arrow), was not tested here.
\textbf{d}, Tools available in each setup (dots; Methods).
\textbf{e}, Files kept for every simulation attempt, so that each revision can be audited.
\textbf{f}, A separate LLM scorer agent checks study-specific qualitative and quantitative criteria against the agent-generated model and returns a score (0--1); shown are the criteria for the study in \textbf{a}. hph, hours post-hatching. Material from Atwell \textit{et al.}\cite{atwell2015} adapted under the \href{https://creativecommons.org/licenses/by/3.0/}{CC BY 3.0} license.}
\label{fig:workflow}
\end{figure*}

Many open questions in biological development, regeneration and cancer require mechanistic models that connect biological observations to hypotheses about cells, forces and signaling (Fig.~\ref{fig:workflow}a).\cite{boutillon2022,vu2019,ilina2020}
Building such a multiscale virtual tissue model is slow, specialist work.\cite{fletcher2022}
A modeler must abstract observations into assumptions, cast them in a mathematical framework, implement them in a simulator such as PhysiCell,\cite{ghaffarizadeh2018} CompuCell3D\cite{swat2012} or Morpheus,\cite{starruss2014} and check the output against expected behavior; each step needs different expertise (Fig.~\ref{fig:workflow}b).

Large language model (LLM) agents call tools, run their own code and use the feedback to repair failed attempts.\cite{schick2023,chen2024}
In biology, agents already query model repositories in natural language,\cite{wehling2025} assemble multiscale PhysiCell models through Model Context Protocol (MCP) servers\cite{ruscone2026} and synthesize literature rules for a natural-language simulation grammar.\cite{johnson2025a,sims2025}
Benchmarks have also begun to test how well LLMs design and analyze experiments on systems-biology models,\cite{duan2025,pomereu2026,doresic2026} while general research agents plan studies and drive laboratory robots.\cite{huang2026a}
Where agents built simulations, however, a human ran them or stayed in the loop,\cite{ruscone2026,chen2026a} and code quality depended on how that human responded.\cite{shafique2026}
Whether an agent can build, run and refine a virtual tissue model from the literature on its own, and what it needs to do so, is therefore unknown.

Here, we pair LLM agents with the widely used modeling and simulation framework Morpheus,\cite{starruss2014} which suits this test for two reasons.
First, each Morpheus model is a single declarative MorpheusML file that can be validated against a schema, so a modeler can check, edit and simulate what the agent built in a graphical user interface (GUI) without reading simulation code.
Second, a curated library of validated reference models lets us test how much examples help the agent.

We termed the agents ``MorpheusAI'' and gave them five tool setups that mimic common ways of using LLMs (Fig.~\ref{fig:workflow}c,d; Methods): web search and Python (Web), as in a chatbot; the Morpheus command-line interface (CLI), as in an agentic harness; the CLI with only the documentation and no web search (Docs), as for in-house software; an agent skill including the reference library (Skill); or the skill paired with an MCP server (MCP).
No human intervened while the agent generated, ran, inspected and revised its models, and every attempt was kept for audit (Fig.~\ref{fig:workflow}e).

\begin{figure*}[t]
\centering
\includegraphics[]{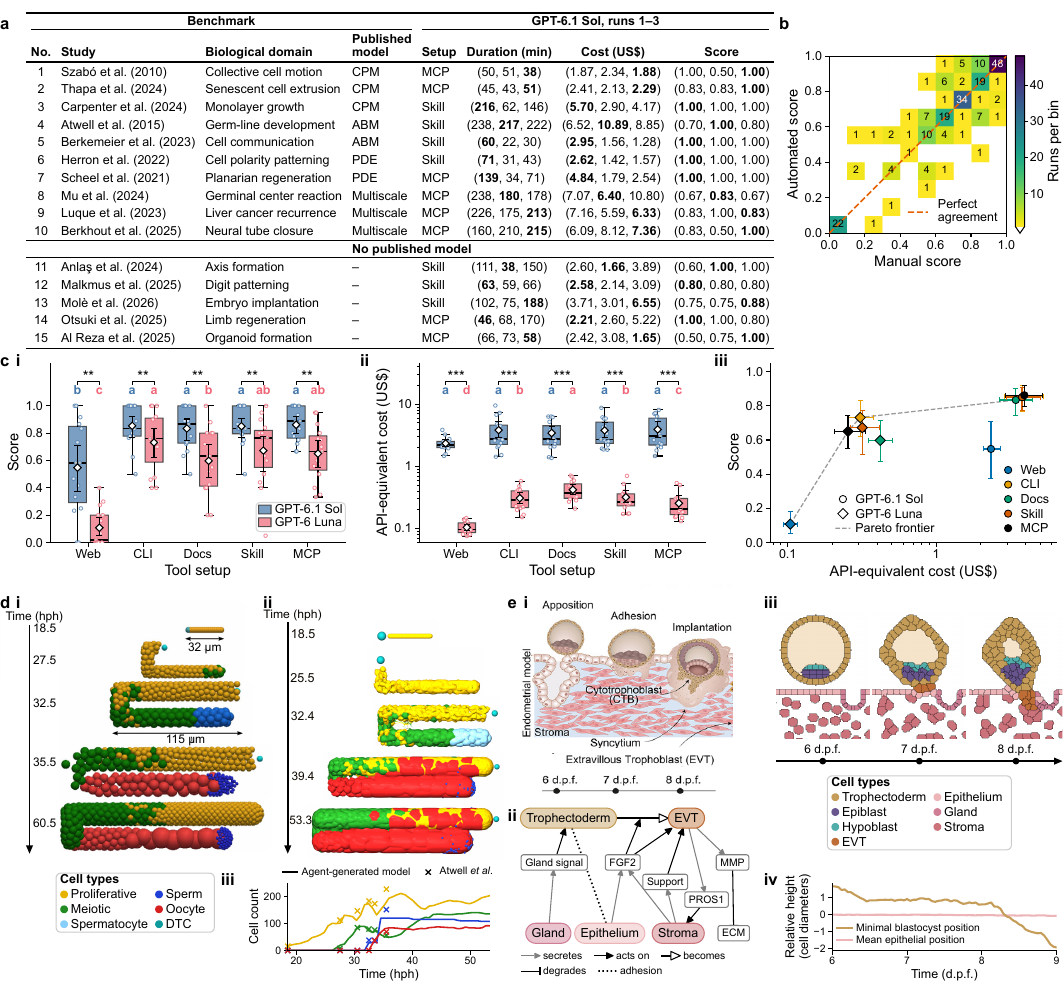}
\caption{\textbf{Benchmark scope and results by MorpheusAI.}
\textbf{a}, The 15 benchmark studies (model formalisms: CPM, cellular Potts; ABM, agent-based; PDE, partial differential equations) and three GPT-6.1 Sol runs per study with the Skill or MCP setup, whichever produced the best-scoring run (ties decided on the simulation outputs; for study 9, see Methods). Bold, the run shown in \textbf{d, e} and Supplementary Figs.~1--15. 
\textbf{b}, Agreement of the automated LLM scorer with manual expert scores for 225 earlier runs (GPT-5.5; Methods); bins 0.1 wide. Dashed line, perfect agreement.
\textbf{c}, Score (\textbf{i}), API-equivalent cost (\textbf{ii}) and both together (\textbf{iii}) for each tool setup with GPT-6.1 Sol and GPT-6 Luna. In \textbf{i} and \textbf{ii}, points are study means over repetitions ($n=15$), boxes show median and quartiles (whiskers, $1.5\times$ interquartile range), and diamonds the mean with its 95\% bootstrap confidence interval (CI); asterisks compare the two LLMs (** $p<0.01$, *** $p<0.001$), and setups of one LLM that share no letter differ ($p<0.05$; tests in Methods). 
\textbf{iii}, Mean score against mean cost per run (log scale); bars, 95\% bootstrap CIs; dashed line, Pareto frontier.
\textbf{d}, Germ-line development in \textit{C.~elegans}. \textbf{i}, simulation snapshots from Atwell \textit{et al}.\cite{atwell2015} (adapted from their Fig.~3). \textbf{ii}, The agent's model (GPT-6.1 Sol, Skill; bold in \textbf{a}) at comparable stages. \textbf{iii}, Germ-cell numbers in the agent's model (lines) and in Atwell \textit{et al}.\cite{atwell2015} (crosses; data from their Table~S2), colored by cell type as in \textbf{i}.
\textbf{e}, Human embryo implantation, a study without a published model; agent's model built by GPT-6.1 Sol (Skill; bold in \textbf{a}).
\textbf{i}, Stages observed by Molè \textit{et al}.\cite{mole2026} (adapted from their graphical abstract). 
\textbf{ii}, Interactions in the agent's model; cell types (colored as in \textbf{iii}), secreted signals and extracellular matrix (white boxes).
\textbf{iii}, Simulation at 6, 7 and 8 d.p.f. \textbf{iv}, Height of the blastocyst's lowest point and mean epithelial height relative to the epithelial surface at 6~d.p.f., in epithelial cell diameters.
CLI, command-line interface; d.p.f., days post-fertilization; DTC, distal tip cell; ECM, extracellular matrix; EVT, extravillous trophoblast; hph, hours post-hatching; MCP, Model Context Protocol.
Material from Atwell \textit{et al}.\cite{atwell2015} is adapted under a \href{https://creativecommons.org/licenses/by/3.0/}{CC BY 3.0} license; material from Molè \textit{et al}.\cite{mole2026} under \href{https://creativecommons.org/licenses/by/4.0/}{CC BY 4.0}.
}
\label{fig:results}
\end{figure*}

We evaluated the agent on 15 studies spanning cell migration, morphogenesis, pattern formation, development, cancer and regeneration (Fig.~\ref{fig:results}a, Supplementary Figs.~1--15, Supplementary Videos~1--15 and Methods).
Ten describe a computational model built in other software or custom code, none in Morpheus.
Two of these models are off-lattice and force-based, a formalism Morpheus lacks, so the agent had to translate the mechanisms.
The other five report only experimental observations, so the agent had to formulate the mechanisms itself.

For each study we defined three to ten weighted qualitative and quantitative criteria (median five; Supplementary Table~1), which the agents never saw; the score is the fraction of the total weight a model met (0--1).
An LLM scorer rated each final simulation against these criteria (Fig.~\ref{fig:workflow}f); on 225 runs of an earlier benchmark round with GPT-5.5, scored with the same criteria, it agreed with manual expert scores (Pearson $r=0.93$; Fig.~\ref{fig:results}b).
Each setup ran with two OpenAI LLMs at reasoning effort ``max'' and a \qty{4}{\hour} budget per run: GPT-6.1 Sol three times per study and the smaller GPT-6 Luna five times.

The agents built models that met most of the criteria (Fig.~\ref{fig:results}c i--iii).
With GPT-6.1 Sol, every setup that could run Morpheus reached a mean score of 0.83--0.86, after \qtyrange{105}{110}{\minute} of active agent time and for US\$3.74 per run on average.
Across these setups with GPT-6.1 Sol, studies without a published model, whose criteria also set numeric targets such as four to six pattern wavelengths in the limb bud (study 12; Methods), scored about as high as those with one (0.83 versus 0.86), suggesting that the agent formulated new mechanisms about as well as it re-implemented published ones.

To get there, the agent had to simulate its model and inspect the output.
With GPT-6.1 Sol, agents that could only write MorpheusML scored 0.55, against 0.83--0.86 for every setup that could run Morpheus; with GPT-6 Luna the gap was wider, 0.11 against 0.60--0.73 (two-sided Wilcoxon signed-rank tests paired by study, Holm-adjusted $p\le0.002$ for every comparison).
Without execution the agents also stopped sooner, after a mean of \qty{58}{\minute} rather than \qtyrange{105}{110}{\minute} with GPT-6.1 Sol and \qty{27}{\minute} rather than \qtyrange{64}{82}{\minute} with GPT-6 Luna ($p\le0.017$), presumably because they had no simulation output to revise against.

The choice of LLM mattered more than the setup.
GPT-6.1 Sol outscored GPT-6 Luna in every setup (0.85 versus 0.66 across the setups that could run Morpheus; $p\le0.006$) at 10.8 times the cost per run (geometric-mean ratio, 95\% CI 8.9--12.7; US\$3.74 versus US\$0.32; Fig.~\ref{fig:results}c).

How much the setup mattered beyond execution depended on the LLM.
With GPT-6.1 Sol, the four setups that could run Morpheus were statistically indistinguishable in score, active agent time and cost (Friedman tests, $p=0.89$, $0.58$ and $0.65$), and the agent reached 0.83 even without web search, with only the MorpheusML documentation and a model template.
Across the four Morpheus-running setups, the smaller GPT-6 Luna was sensitive to its tools ($p=0.011$): it scored lowest without web search (Docs; 0.60), best with the plain CLI and web access (0.73), and no higher with the reference library (0.65--0.67).
Among these setups, Docs also cost most and took longest (US\$0.42 versus US\$0.25--0.32; \qty{82}{\minute} versus \qtyrange{64}{67}{\minute}), whereas an MCP server made it cheapest (US\$0.25 versus US\$0.31--0.42 per run; Methods).

The agent adapted models to what Morpheus can represent.
For germ-line development in the nematode \textit{Caenorhabditis elegans},\cite{atwell2015} it translated a three-dimensional off-lattice model into a cellular Potts model in a U-shaped gonad domain it built itself, with proliferative, meiotic, sperm and oogenic cells (Fig.~\ref{fig:results}d).
Although the agent had to parametrize the new formalism from scratch, its model reproduced the four germ-cell populations, with counts close to the published ones at every published time point it reached (\qtyrange{18.5}{35.5}{\hour} post-hatching; Fig.~\ref{fig:results}d iii).
The simulation ended at about \qty{53}{\hour} rather than the requested \qty{87.5}{\hour} because the benchmark caps each simulation at \qty{60}{\minute} of run time (Methods).
In another study, the agent's model went beyond literal re-implementation: it formed the target rosette pattern from random initial noise, whereas the published simulations needed a weak rosette-shaped pre-pattern (Supplementary Fig.~6).\cite{herron2022}

The agent also formulated mechanisms from scratch for studies without a published model.
For human embryo implantation in an endometrial model,\cite{mole2026} it formalized the reported ligand--receptor interactions (EFNA1--EPHA2 and NECTIN3--CADM1) as trophectoderm--epithelium adhesion and let trophectoderm differentiate into extravillous trophoblast after 24 h of epithelial contact in the presence of FGF2 and stromal--gland signaling; the trophoblast then secreted MMP to degrade the extracellular matrix (Fig.~\ref{fig:results}e ii).
In the simulation, the three stages of apposition, adhesion and implantation are discernible, and the blastocyst crossed the epithelial surface after 8 days post-fertilization, as in the \textit{in vitro} model, where embryos implanted by 8 and had penetrated the epithelium by 10 days post-fertilization (Fig.~\ref{fig:results}e i, iii, iv).

LLM agents that can run, inspect and revise their own simulations thus build virtual tissue models from the literature without human help, within hours and for a few dollars per model.
Execution feedback is known to help LLMs write code;\cite{chen2024} here it raised the score for emergent multicellular behavior by 0.29--0.31 with GPT-6.1 Sol and 0.49--0.62 with GPT-6 Luna.
We expect simulators that run declarative, schema-checked models without recompilation to suit agents particularly well, because every revision is then a short, validated model file whose simulation the agent can run and inspect within minutes.

Our criteria test three to ten features per study (median five), and four studies scored 1.0 on every Morpheus-running GPT-6.1 Sol run, so harder criteria and new studies will be needed to keep the benchmark discriminative for future LLMs; it still separates the two LLMs and setups with and without execution.
Although the simulations are stochastic, we scored each model on a single run and chose numeric criteria with ranges wide enough that the score does not depend on the seed: across ten random seeds, the agent's Notch--Delta model (study 5) reached high-Delta cell fractions of 0.18--0.20 (Supplementary Fig.~5d) and passed every run.
Similarly, testing a single simulation platform and LLM provider limits cross-platform generality, yet suffices to test whether the autonomous loop works and which ingredients it needs.

AI scientists that generate hypotheses, analyze data or automate research end to end promise to accelerate discovery,\cite{lu2026,gottweis2026,ghareeb2026} but audits of their trace logs and code have revealed data leakage, metric misuse and post-hoc selection bias.\cite{luo2026}
We therefore kept generation and evaluation apart: the agent's deliverable is a single model file in the standardized MorpheusML format, which we re-execute in Morpheus without the agent and score against criteria it never saw, using an LLM scorer validated against human expert modelers.

Where no model existed, the agents' models state their mechanistic hypotheses explicitly in MorpheusML, so they can be simulated, compared with alternatives and refined \textit{in silico} before a costly wet-lab experiment.
Such agents could make mechanistic modeling routine in biology, even for researchers without training in mathematics or programming.

\section*{Methods}

\subsection*{Morpheus and MorpheusML}
Morpheus (\url{https://morpheus.gitlab.io}) is an open-source platform for multicellular modeling, and MorpheusML (\url{https://doi.org/10.25504/FAIRsharing.78b6a6}) is its declarative XML format for specifying individual simulation experiments.\cite{starruss2014}
An entire model specification, including space, time, cell types, cell behaviors, cellular Potts dynamics, ordinary and partial differential equations and output analysis, is stored in a single file that can be validated against the MorpheusML schema, simulated with Morpheus CLI, reopened in the Morpheus GUI, inspected through parameter dialogues and interactive model graphs, and extended with imported SBML modules.\cite{hucka2003}

\subsection*{Agent harness}
Each run used an OpenAI Codex agent in a file-system sandbox: the agent could read and write only its own workspace, and could otherwise read only the minimal system files needed to run commands, a locked Python environment and, where its setup provided them, the Morpheus executable and skill.
All other files, including other runs, the scoring criteria and the code running the benchmark suite, were inaccessible.
The agent received the source PDF, its extracted text and a short paper-specific prompt naming the target observables and simulation time window.
Supplementary Note~1 contains the complete prompt for the agent that produced Fig.~\ref{fig:results}e (Study 13), and all prompts are in the code repository (Code availability).
A common prompt asked it to preserve the reported mechanisms, parameters, scales and cell counts as closely as Morpheus allows, and to avoid shortcuts that reproduce a figure without the underlying mechanism, such as pre-placing future cells.
The deliverable was a single \texttt{model.xml} file.
Every simulation ran in a fresh attempt folder holding the exact model, logs and outputs, so that each revision can be audited.
The agent could view pages of the source PDF and its own output images and data.

\subsection*{Agent setups}
The five setups differ only in the tools exposed to the agent; task, prompt, time budget and scoring were identical (Fig.~\ref{fig:workflow}d, Supplementary Note~1).
\begin{itemize}
\item \textbf{Setup 1} (Web): web search and a Python environment, emulating a chatbot such as ChatGPT. The agent writes MorpheusML but cannot execute Morpheus and therefore never sees a simulation result; it can only check the XML statically.
\item \textbf{Setup 2} (CLI): Setup 1 plus the Morpheus command-line interface, mimicking a general agentic harness such as Codex or Claude Code without domain-specific instructions, so that the agent can run its model and inspect logs, CSV files and images. No Morpheus documentation or examples are provided beyond what the agent finds online.
\item \textbf{Setup 3} (Docs): the Morpheus CLI and Python without internet access, simulating proprietary or in-house software for which no examples are available. The only Morpheus-specific material is the MorpheusML reference documentation and a minimal model template; no example models are provided.
\item \textbf{Setup 4} (Skill): Setup 2 plus a Morpheus agent skill file: a packaged set of instructions with the MorpheusML documentation and a curated library of 43 didactic and validated example models grouped by formalism (cellular Potts, ODE, PDE, multiscale and miscellaneous).
The skill also pointed the agent to the Morpheus source code (\url{https://gitlab.com/morpheus.lab/morpheus}) and the Morpheus model repository (\url{https://gitlab.com/morpheus.lab/model-repo}) of more than 100 models.
\item \textbf{Setup 5} (MCP): Setup 4 plus an MCP server, a second way to specialize an agentic harness, which exposes Morpheus functionality as typed tools: paper ingestion and page rendering, reference lookup, XML writing with schema validation, simulation execution, run summaries and output image sampling.
\end{itemize}

\subsection*{Benchmark}
While an agent can be tasked with custom problem statements and data, we standardized the task input to the content of a single scientific paper.
The benchmark comprised 15 studies (Fig.~\ref{fig:results}a).
Thirteen were published in 2021 or later.
Ten (studies 1--10) describe a computational model: collective migration,\cite{szabo2010} ovarian cancer colonization,\cite{thapa2023} epithelial proliferation,\cite{carpenter2024} germ-line dynamics,\cite{atwell2015} Notch--Delta signaling,\cite{berkemeier2023} phagocytosis,\cite{herron2022} planarian regeneration,\cite{scheel2021} the germinal center response,\cite{mu2024} liver regeneration with tumor growth\cite{luque2023} and neural tube closure defects.\cite{berkhout2025}
None of these models was implemented in Morpheus; the germ-line\cite{atwell2015} and liver\cite{luque2023} models are off-lattice, force-based cell models, a formalism Morpheus does not provide.
The remaining five (studies 11--15) report experimental observations without a computational model, so the agent had to formulate the mechanism itself: gastruloid axis formation,\cite{anlass2024} digit patterning,\cite{malkmus2025} embryo implantation,\cite{mole2026} limb regeneration\cite{otsuki2025} and liver organoid zonation.\cite{alreza2025}
With its criteria, the benchmark can also test other agentic modeling frameworks:\cite{ruscone2026,chen2026a} even after excluding the studies whose published model used a given framework, more than ten problems remain new to it.
Because the agents' final models and our scoring criteria are now public, future runs could retrieve them instead of building their own models; reruns should therefore block these sources, audit the transcripts for retrieval and add studies published after the LLMs' training cutoff.

\subsection*{Time budget and run protocol}
Each run received one continuous agent execution with a wall-clock budget of \qty{4}{\hour}, measured from the start of the agent process.
During the final \qty{10}{\minute} the launcher refused new simulations so that the agent could inspect its outputs and finalize; individual simulations were capped at \qty{60}{\minute} or the time remaining before that reserve, whichever was shorter.
The agent could query the remaining time at any point.
There was no limit on the number of tool calls, turns or simulation attempts, and no human intervened during a run.
Failures of the infrastructure, for example a sandbox failure before the agent started, an exhausted service quota, or LLM-service provider issues, were rerun from scratch with an empty workspace and the full budget; time-outs, invalid XML and failed simulations counted as benchmark outcomes and were not rerun.
We ran all five setups on all 15 studies with two OpenAI LLMs at reasoning effort \texttt{max}, accessed through the Codex SDK and command-line interface (version 0.160.0): GPT-6.1 Sol (model identifier \texttt{gpt-6.1-sol}) with three repetitions (225 runs, 7--14 October 2026) and the smaller GPT-6 Luna (\texttt{gpt-6-luna}) with five repetitions (375 runs, 3--4 October 2026). Simulations ran in Morpheus 2.4.1 through its command-line interface.
Median active agent time, the wall-clock time of the agent process, was \qty{70}{\minute} for GPT-6.1 Sol and \qty{44}{\minute} for GPT-6 Luna; 4 of the 225 Sol runs and none of the 375 Luna runs reached the time limit.
The main text reports means, which exceed these medians because active times are right-skewed.

\subsection*{Scoring}
After the agent finished, we automatically re-executed the final \texttt{model.xml} in Morpheus without the agent; this canonical attempt is the basis for scoring.
Generation and evaluation were kept separate: the agent never saw the scoring criteria or their weights, and the scorer assessed the canonical attempt; it could also read the agent's workspace, but not change it.
For each study we defined a set of binary criteria that a faithful model should satisfy, with three to ten criteria per study (median five; Supplementary Table~1 and the code repository).
For studies 1--10 the criteria are split into qualitative criteria (for example the emergence of a specific spatial pattern) and quantitative criteria (for example a mean cell speed within a stated range), with weights chosen so that both categories carry equal total weight.
For studies 11--15, which report no model output to compare against, all criteria carry equal weight; they combine spatial target patterns with numeric thresholds (for example a two- to threefold increase in cell number or a target wavelength of a pattern).
A separate scoring agent (GPT-6 Luna, reasoning effort \texttt{high}) received the criteria, the canonical \texttt{model.xml} and its outputs, including time-ordered output images, in a read-only workspace without network access, and returned a verdict, the observed value and the evidence for every criterion.
The score is the passed weight divided by the total weight (0--1).
Runs whose canonical simulation failed or did not have scoreable output received a score of 0 without invoking the scorer. In study 9, the scorer passed a criterion requiring a ratio between 10 and 1,000 for two GPT-6.1 Sol runs whose ratios lay outside that range (1,756 with MCP, repetition 2; 2,861 with CLI, repetition 1). We report the scorer's raw output throughout, but excluded repetition 2 when selecting the bolded representative run for study 9 in Fig.~\ref{fig:results}a (and Supplementary Fig.~9).
To validate the automated scorer, three of the authors (P.C., A.S. and L.B.) scored the 225 runs of an earlier benchmark round (GPT-5.5, reasoning effort \texttt{xhigh}, three repetitions) by hand with the same criteria; each run was scored by one of them, a second author re-examined doubtful cases, and the reported score is their consensus.
Automated and manual scores agreed exactly in 139 of 225 runs (62\%), with a mean absolute difference of 0.059, a mean bias of $+0.012$ (automated minus manual), Pearson $r=0.93$ and an ordinary least-squares fit of automated on manual score with slope 0.95, intercept 0.04 and $R^2=0.86$ (Fig.~\ref{fig:results}b).
During model generation, the modeling agent also rated each of its own attempts on a 0--8 reproduction rubric; these self-scores guided its iterative model refinement but were not used for evaluation.

\subsection*{Statistics}
For the results in Fig.~\ref{fig:results}c, the unit of replication is the study ($n=15$).
Each study was run several times with every setup, so we first averaged these repetitions, which gives one score per study and setup; setups were then compared on these paired study-level values.
We report means across studies with 95\% bias-corrected and accelerated (BCa) bootstrap confidence intervals, resampling studies (10,000 resamples).
Setups were compared with two-sided Wilcoxon signed-rank tests paired by study, keeping zero differences in the ranking (Pratt) and computing exact $p$-values from the sign-flip distribution, which remain valid with tied scores. 
We compared the two LLMs within each setup, and each setup that could run Morpheus with the Web setup.
Within each LLM, a Friedman test across the four setups that could run Morpheus was followed by tests of all pairs of these setups.
$p$-values were Holm-adjusted within each family of comparisons; the letters in Fig.~\ref{fig:results}c use Holm adjustment across all ten pairs of setups within an LLM.
Cost per run and active agent time were compared in the same way.
Both were right-skewed and differed between setups by roughly constant factors rather than constant amounts, so we compared them on a log scale: the Wilcoxon test was applied to the paired log ratios (one per study), and effects are reported as geometric-mean ratios, the exponential of the mean log ratio across the 15 studies, with 95\% BCa bootstrap confidence intervals.
Among the four Morpheus-running setups for GPT-6 Luna, Docs cost more than each other setup ($p\le0.002$ for each pair) and took longer than CLI and MCP ($p<0.001$), whereas MCP cost less than each other Morpheus-running setup ($p\le0.025$ for each pair); Web cost least overall because runs stopped early without execution.

\subsection*{Cost and token usage}
Input, cached input, output and reasoning tokens were recorded for every run from the Codex event stream, and API cost was computed from these counts using OpenAI's standard-tier list prices per million tokens at the time of the runs (GPT-6.1 Sol: US\$2.00 input, US\$0.10 cached input, US\$10.00 output; GPT-6 Luna: US\$0.10, US\$0.01 and US\$0.50), with reasoning tokens billed as output.
Mean cost per run across the four setups that could run Morpheus was US\$3.74 for GPT-6.1 Sol and US\$0.32 for GPT-6 Luna.

\subsection*{Use of generative AI}
The LLM agents under study generated all benchmark models and simulation outputs: GPT-6.1 Sol and GPT-6 Luna through OpenAI Codex, and GPT-5.5 for the earlier benchmark round in Fig.~\ref{fig:results}b; GPT-6 Luna also scored the runs (Scoring).
In addition, the authors used OpenAI Codex and Claude Code (Anthropic) as assistants.
Both helped write the benchmark runner, the MCP server and the Morpheus skill.
Claude Code helped write the analysis, statistics and figure code, check references against their sources, and draft and edit passages of the manuscript.
The authors reviewed and verified all AI-assisted code and text.

\subsection*{Data availability}
The \texttt{figures/} folder of the code repository and its Zenodo archive (Code availability) contains the data underlying Fig.~\ref{fig:results} and Supplementary Figs.~1--15, and scripts that rebuild every figure from them.
The final \texttt{model.xml} file of each benchmark study, used for the simulation snapshots in Supplementary Figs.~1--15 and Supplementary Videos~1--15, is provided as Supplementary Data.
These results and figures can be reproduced from the provided models by running Morpheus 2.4.1 through its command-line interface. 
Opening these models in the Morpheus 2.4.1 GUI replaces certain attributes (such as hex-coded colors) with defaults; the GUI Fixboard lists each replacement for manual adjustment (for example, to the closest named color).
The raw outputs of all benchmark runs (GPT-6.1 Sol, GPT-6 Luna and the earlier GPT-5.5 benchmark) for the 11 studies published under a Creative Commons Attribution license, with the agent workspaces and transcripts, every model version, the Morpheus runs and the scorer outputs, are available on Zenodo under a CC BY 4.0 license at \url{https://doi.org/10.5281/zenodo.24535542}.
The run folders of the other four studies are not public because they could leak their copyrighted PDFs; we provide them upon reasonable request.

\subsection*{Code availability}
The benchmark runner, the MCP server, the Morpheus skill, the agent prompts, the scoring criteria and the analysis scripts are available at \url{https://github.com/sisyga/morpheus-agent-benchmark} under the Apache License 2.0; it is archived on Zenodo at \url{https://doi.org/10.5281/zenodo.23697418}.
To use the workflow for other modeling tasks, install the Morpheus skill from \url{https://github.com/sisyga/morpheus-skills}, which explains its setup.

\subsection*{Author contributions}
P.C.: conceptualization, methodology, software, validation, formal analysis, investigation, data curation, visualization, writing (original draft, review and editing).
A.S.: validation (manual scoring).
J.B.M. and J.S.: software (Morpheus).
D.J.: resources (Morpheus model repository).
L.B.: conceptualization, investigation (CompuCell3D simulations for Supplementary Fig.~10), validation (manual scoring), supervision, funding acquisition, writing (original draft, review and editing).
S.S.: conceptualization, methodology, software, validation, formal analysis, investigation, data curation, visualization, writing (original draft, review and editing), supervision.
All authors read and approved the manuscript.

\subsection*{Competing interests}
The authors declare no competing interests.

\subsection*{Funding}
This work has received funding from the FOSTER program of TUD Dresden University of Technology to P.C., A.S. and J.B.M. (grant no. ZE3-ZIH).
P.C. is funded by the German National High Performance Computing Association (NHR-Verein e.V.).
We acknowledge funding from the German Federal Ministry of Research, Technology and Space (BMFTR) within the LiSyM-Cancer-II network (grant no. 031L0315A) to L.B. and from the OSCARS project to L.B., which has received funding from the European Commission’s Horizon Europe Research and Innovation programme (grant no. 101129751). 
This work has further received funding from the Volkswagen Foundation to L.B. within the program Impetus for the German Research System: Pioneer Projects (grant no. 0200092-00).
S.S. is funded by the European Union (ERC, subLethal, 101054921, \url{https://erc.europa.eu/}). Views and opinions expressed are however those of the authors only and do not necessarily reflect those of the European Union or the European Research Council Executive Agency. Neither the European Union nor the granting authority can be held responsible for them.
S.S. was supported by the Graduate Academy of TUD Dresden University of Technology, funded by the Excellence Initiative of the German Federal and State Governments.

\subsection*{Acknowledgements}
We are grateful to the Center for Information Services and High Performance Computing (ZIH) of TUD Dresden University of Technology and NHR@TUD for the provision of computing power.

\bibliographystyle{naturemag}
\bibliography{preprint_v2_references}

\end{document}